\documentclass[prd, twocolumn, superscriptaddress, nofootinbib, preprintnumbers]{revtex4-2}

\usepackage[bottom]{footmisc}
\usepackage{hyperref}
\usepackage[normalem]{ulem}
\usepackage[utf8]{inputenc}
\usepackage{amsmath}
\usepackage{mathrsfs}
\usepackage{graphicx}
\usepackage{braket}
\usepackage{color}
\usepackage[utf8]{inputenc}
\usepackage{etoolbox}

\newcommand{\lb}{\left (}
\newcommand{\rb}{\right )}

\newcommand{\ob}{\Omega_{\mathrm{b}}}
\newcommand{\om}{\Omega_{\mathrm{m}}}
\newcommand{\oL}{\Omega_{\Lambda}}
\newcommand{\dl}{\delta_L}
\newcommand{\mpc}{\mathrm{Mpc}}
\newcommand{\msun}{\mathrm{M}_{\odot}}
\newcommand{\lcdm}{\Lambda\mathrm{CDM}}
\newcommand\lsim{\mathrel{\rlap{\lower4pt\hbox{\hskip1pt$\sim$}}\raise1pt\hbox{$<$}}}
\newcommand\gsim{\mathrel{\rlap{\lower4pt\hbox{\hskip1pt$\sim$}}\raise1pt\hbox{$>$}}}
\newcommand{\graphic}[2]{\includegraphics[width=#2\linewidth, type=pdf,ext=.pdf,read=.pdf]{#1}}

\newcommand{\dee}{\mathrm{d}}
\newcommand{\brho}{\bar{\rho}}

\newcommand{\pdf}{\mathscr{P}}
\newcommand{\V}{\mathcal{V}}
\newcommand{\su}{\mathrm{su}}

\begin{document}
	
	\preprint{YITP-SB-2020-40}
	\title{Position-dependent Voronoi probability distribution functions for matter and halos}
	\author{Drew Jamieson and Marilena Loverde \\{\it{\small  C.N. Yang Institute for Theoretical Physics, Department of Physics \& Astronomy, Stony Brook University, Stony Brook, New York, 11794, USA}}}
	
	\begin{abstract}
		We measure the Voronoi density probability distribution function (PDF) for both dark matter and halos in N-body simulations. For the dark matter, Voronoi densities represent the matter density field smoothed on a uniform mass scale, which approximates the Lagrangian density field. For halos, the Voronoi densities contain information about the local environment of each halo. We measure the halo virial masses, the total amount of dark matter within each halo Voronoi cell, and the halo Voronoi cell volumes, and we show how halo abundances depend on these three quantities. We then study the position-dependent Voronoi density PDF, measured within finite subregions of the Universe, using separate universe simulations. We demonstrate that the spatial variation of the position-dependent PDF is due to large-scale density fluctuations, indicating that the position-dependent PDF is a biased tracer of large-scale structure. We measure this bias for the dark matter, and interpret it as the bias of regions of the Lagrangian density field that are selected based on density. For the halos, this bias can be interpreted as a form of assembly bias. We present the mapping from late-time to early-time Voronoi density for each simulation dark matter particle, which is highly stochastic. We compare the median of this stochastic map with spherical collapse calculations and discuss challenges involved in modeling the evolution of the density field on these scales.
	\end{abstract}
	
	\maketitle
	
	\section{Introduction}
	\label{sec:intro}
	
		\begin{figure*}
			\graphic{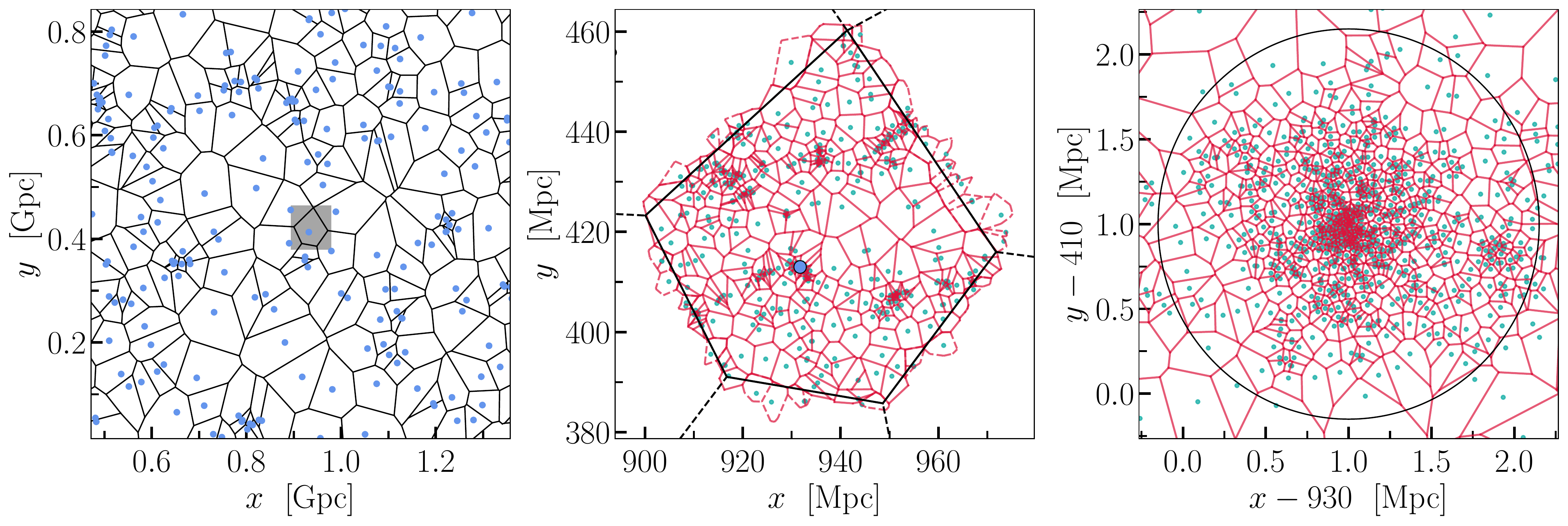}{1.}
			\caption{Left: 2D Voronoi tessellation of simulation halos with mass $M > 10^{13}~\mathrm{M}_\odot$  in a slice of width $0.4~\mathrm{Mpc}$ taken at redshift $z=0.0$. The halos are indicated by blue dots, and the boundaries of their Voronoi cells are the black lines. The shaded square indicates the area that is shown in the middle plot. Middle: 2D Voronoi tessellation of the dark matter particles within a halo's Voronoi cell. The particles are indicated by green dots and their cell walls are the red lines. The dashed red lines indicate dark matter cells that overlap with the halo's cell while the dark matter particle lies outside of the halo's Voronoi region. The halo has mass $M=8.7\times 10^{13}~\mathrm{M}_\odot$ and its position is indicated by the blue circle with radius corresponding to the halo's virial radius $R=1.1~\mathrm{Mpc}$. Right: 2D Voronoi tessellation of the dark matter within the halo's virial radius.}
			\label{fig:voro}
		\end{figure*}
	
		The spatial distributions of matter and galaxies contain an abundance of information about the contents, evolution, and origins of the Universe. With upcoming large-scale structure (LSS) surveys \cite{Ivezic_2008, Laureijs_2011, Spergel_2015, Aghamousa_2016}, we will improve upon the already impressive measurements of cosmological parameters from CMB data \cite{Aghanim:2018eyx}, and possibly detect evidence of new physics beyond $\lcdm$, the standard model of cosmology. The physical processes of structure formation involve nonlinear gravitational clustering, which must be theoretically understood and accurately modeled in order to extract optimal and rigorous constraints from observational data. For this reason, the theory and phenomenology of LSS formation continues to be an important area of research in cosmology. 
	
		The first step of any LSS analysis is to decide how to smooth the density field. In the Eulerian description, the density field is smoothed on a fixed comoving length scale, so that density fluctuations represent mass fluctuations. In the Lagrangian description, the density field is smoothed into equal-mass fluid cells, which flow under the competing influence of cosmic expansion and gravitational clustering. Since the fluid cells all contain the same amount of mass, the density fluctuations of the Lagrangian field represent volume fluctuations.  
		
		These two approaches have complementary strengths and weaknesses. The fixed comoving length scale imposed under Eulerian smoothing allows the level of nonlinearity to be controlled, by averaging out the highly nonlinear small scales. Then the Eulerian density field can be described in the linear and quasilinear regimes using the effective field theory of LSS, which can provide tighter parameter constraints than analyses with standard perturbation theory \cite{DAmico:2019fhj, Colas:2019ret, Ivanov:2019hqk, Philcox:2020vvt}. The trade-off is that many structural features of the cosmic web, which contain cosmological information, may be washed out by this smoothing. The mass-weighted Lagrangian smoothing provides a more detailed description of the filamental structures in the cosmic web \cite{vandeWeygaert:1993ev, vandeWeygaert:2007ze, AragonCalvo:2007mk}, while losing some control over the level of nonlinearity.
	
		In this work, we study the one-point statistics of the Lagrangian matter density field, smoothed on a uniform mass scale through the use of Voronoi tessellation in N-body simulations. Voronoi tessellation only achieves an approximation of the true Lagrangian field, since it neglects overlap of fluid cells due to multistreaming effects \cite{ Shandarin:2011jv}. However, the tessellation allows us to probe the distribution of matter deep into the nonlinear regime, and test our understanding of structure formation on these scales. We also study the one-point statistics of the halo number density field under Voronoi tessellation, which probes the local environment of halos and their nearest neighbors \cite{Cooper:2005ci}. 
		
		The position-dependent one-point statistics, measured within finite regions of the Universe, are spatially modulated by the long-wavelength modes of the density field, which couple to small-scale modes due to nonlinear gravitational evolution. This is analogous to the spatial modulations of halo and galaxy abundances, which are parameterized bias models. The concept of cosmic bias can be generalized to any local observable that is affected by the nonlinearity of gravitational clustering. Previously, the generalized bias of the one-point statistics of the Eulerian density field was measured in N-body simulations and accurately modeled \cite{Jamieson:2020wxf}. In this work we introduce the generalized bias associated with the one-point statistics of the matter and halo density fields smoothed using Voronoi tessellation.
		
		The outline of this paper is as follows. In Sec. \ref{sec:vor}, we define the Voronoi density PDF and the locally measured, position-dependent Voronoi density PDF. In this section, we also present the dependence of halo abundances on their virial mass, the total mass in their Voronoi cell, and their Voronoi volume. We summarize the details of our N-body simulations in Sec. \ref{sec:sim}. Results for the globally defined Voronoi density PDFs in both dark matter and halos are presented in Sec. \ref{sec:gvp}. We review the separate universe formalism in Sec. \ref{ssec:suf}, and  in Sec. \ref{ssec:sur} we present our separate universe determinations of the position-dependent Voronoi PDF for simulation dark matter and halos. We validate our separate universe results using global universe simulations in Sec. \ref{ssec:guv}.  In Sec. \ref{sec:mod}, we discuss the challenges of modeling the matter Voronoi statistics due to stochasticity and nonlinearity in the evolution of the dark matter Voronoi cells, and study the sensitivity of the position-dependent matter PDF to the simulation resolution.  In Sec. \ref{sec:con} we conclude.  
		
		\section{Voronoi densities}
		\label{sec:vor}
		
			LSS surveys provide catalogs of objects that represent the most dense features of the cosmic web. A typical strategy for making cosmological inferences using this data is to construct the number density field of catalog objects and measure its N-point statistics \cite{Gil-Marin:2014sta, Gil-Marin:2014baa, Gil-Marin:2016wya, Slepian:2016kfz, Abbott:2017wau, Pearson:2017wtw, Gil-Marin:2020bct}. This involves smoothing the discrete catalog objects over some window function. Typical choices for the smoothing window function include a spherical top-hat of fixed radius \cite{Yang:2010qs, Bel:2013csa, Hurtado_Gil_2017}, or smoothing B-spline interpolation on a comoving Cartesian grid \cite{1988csup.book.....H, Jing:2004fq}. Both of these involve cells of fixed shape and size, and so they probe the abundance fluctuations of catalog objects and mass fluctuations within each cell. This fixed-volume smoothing gives an estimation of the Eulerian density field.
			
			A complementary approach is to fix the amount of mass, or the number of objects that each cell contains and allow the shapes and sizes of the cells to vary. In the latter case, this gives an equal-mass smoothed estimation of the Lagrangian density field.  One way of achieving this is through Voronoi tessellation, which provides an unambiguous assignment of volume and cell shape to each discrete object in a catalog. Voronoi tessellation has been employed in several contexts for cosmological simulations and observational data analysis, including void finders \cite{Neyrinck:2007gy, Sutter:2014haa}, and cosmic web structure analysis \cite{Bernardeau:1995en, vandeWeygaert:2007ze, Shandarin:2011jv, Neyrinck:2012bf, Paranjape:2020wuc, Chan:2020fnj}.
						
			\subsection{Tessellation}
			\label{ssec:tess}
			
				\begin{figure*}
					\graphic{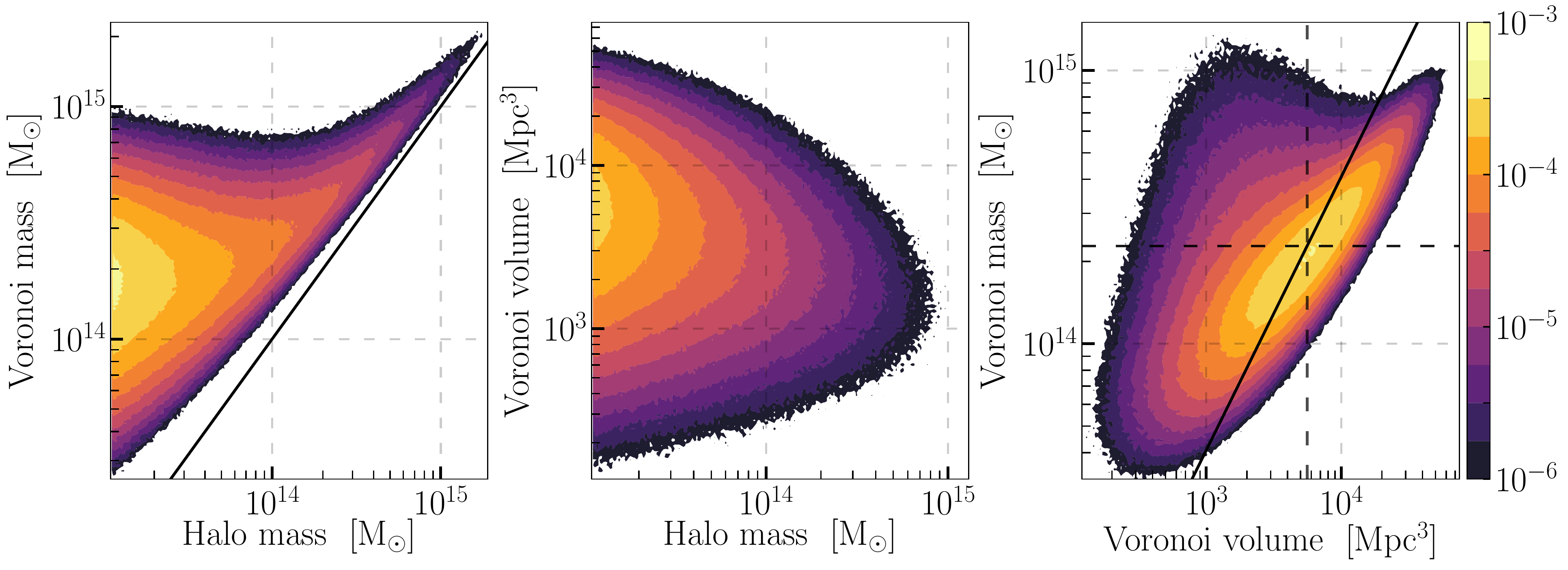}{1.}
					\caption{Left: Number density of halos measured in a 2D histogram of the halo virial mass and the total mass within a halo's Voronoi cell, the color map indicates the value of $\partial^2  \bar{n} / \partial \log M / \partial \log M_\V$, where $ \bar{n}$ is the halo number density in units of $\mpc^{-3}h^3$, $M$ is the halo virial mass, and $M_\V$ is the total mass within the halo's Voronoi cell. The black line corresponds to $M = M_\V$, when all mass within a halo's Voronoi cell equals the mass inside its virial radius. Middle: Number density of halos measured in a 2D histogram of the halo virial mass and the halo Voronoi volume, $\partial^2  \bar{n} / \partial \log M / \partial \log \V$, where $\V$ is the halo's Voronoi volume. Right: Number density of halos measured in a 2D histogram of Voronoi volume and total mass within a halo's cell, $\partial^2  \bar{n} / \partial \log \V / \partial \log M_\V$ . The dashed lines correspond to the mean values of Voronoi volume and total cell mass. The solid line represents mean density cells, with $M_\V=\brho\,\V$. The histograms are binned uniformly in $\log$ of mass and $\log$ of volume, using 200 bins within the ranges of the plot axes, averaged over 20 simulations in $1$~Gpc boxes. The halo catalogs were taken at redshift $z=0.0$ with a lower virial mass threshold $M>10^{13}\ \msun$, and had an average of $5.2\times 10^5$ halos per simulation.}
					\label{fig:voroprops}
				\end{figure*}
			
				Voronoi tessellation is a method for partitioning a region of space around a catalog of discrete, point-like objects. Each object is assigned a Voronoi cell such that the space within an object's cell is nearer to it than to any other object in the catalog. A Voronoi cell is bounded by a set of planes, or faces, that form a polyhedron completely enclosing the cell's volume. The faces meet at edges, three of which meet at a vertex\footnote{In three dimensions it is exceedingly rare for more than three edges to meet at a single vertex, even computationally within floating point accuracy.}. When two cells share a common face, they are said to be neighbors. A cell's shape and size encode information about its distribution of immediate neighbors. There is an enormous amount of information within a Voronoi tessellation, including the statistical distributions of volumes, faces, edges, vertices, and spatial correlations among all of these quantities. In this work, we will focus exclusively on the cell volumes as an estimator of density. 
			
				Two dimensional examples of Voronoi tessellations from a slice through one of our N-body simulations are shown in Fig.~\ref{fig:voro}. The plot on the left is a tessellation around a catalog of halos with mass greater than $10^{13}~\mathrm{\msun}$. One of the halos is highlighted by a gray square, and its Voronoi cell is plotted in greater detail in the middle panel, along with the tessellation of matter particles from the N-body simulation. The halo's Voronoi cell is much larger than its virial radius, and contains complex substructures of clustered matter. The plot on the right shows the Voronoi tessellation of the dark matter particles within the halo's virial radius.
			
				The additional mass in the halo's cell is clustered, and some of the overdensities may be unresolved halos with masses below the threshold of our catalog. By lowering the  mass threshold, some of these low-mass objects will be included in the catalog, and the Voronoi cell of the original halo will be significantly altered, dividing into smaller cells around the newly resolved halos. The halo's Voronoi volume is also not equal to the sum of volumes of all the dark matter particles within its cell. Some contributions to the volume come from particles that lie outside the cell, and some particles within the cell have volumes that extend to its exterior.

				The halo Voronoi properties are plotted as 2D histograms in Fig.~\ref{fig:voroprops}. The plot on the left is the differential, conditional halo mass function,
				\begin{align}
				\frac{\partial^2 \bar{n}(M, M_\V)}{\partial \log M \partial \log M_\V} \, ,
				\end{align}
				where $ \bar{n}(M, M_\V)$ is the cumulative halo number density, for halos with virial mass greater than $M$, and total mass within their Voronoi cells greater than $M_\V$. The middle plot is the same, but for Voronoi volume $\V$ instead of total cell mass,
				\begin{align}
					\frac{\partial^2  \bar{n}(M, \V)}{\partial \log M \partial \log \V} \, .
				\end{align}
				The right plot is,
				\begin{align}
					\frac{\partial^2  \bar{n}(M_\V, \V)}{\partial \log M_\V \partial \log \V} \, .
				\end{align}				
				The color bar on Fig.~\ref{fig:voroprops} shows the values of differential number density in units of $\mpc^{-3}h^3$.
				
				 The low-mass halos have a wide range of masses within their Voronoi cells, and a corresponding wide range of cell volumes. This demonstrates that low-mass halos are found in diverse environments, from strongly clustered, small volume, low-mass cells, to strongly anti-clustered, large volume, high-mass cells. The lowest mass halos, with $M\sim 10^{13}$, are very weakly clustered, in the sense that their distribution is peaked near the mean volume per halo ($5.7\times 10^3~\mpc^{3}$) in the catalog, which is expected for a Poisson sample of points. For high-mass halos with $M > 10^{14}~\mathrm{M}_\odot$, the total mass in a halo's Voronoi cell is much more narrowly peaked near and above the virial mass, which bounds the total mass from below. These halos also have a narrower distribution of Voronoi volumes, which peaks 0.2--0.5 times below the mean volume per halo, indicating their strong clustering.
				
				In the right-most plot of Fig.~\ref{fig:voroprops}, the solid line corresponds to $M_\V = \brho\, \V$, indicating halos with total Voronoi mass density within their cells equal to $\brho$, the mean density of the Universe. Points above (below) this line are overdense (underdense) cells. For unclustered, Poisson distributed matter, the histogram would be symmetric around this line, whereas the peak of the distribution follows a curve that is tilted downwards with respect to this line. On the other hand, for Poisson distributed halos, the distribution of volumes is much narrower and the distribution of total cell masses is wider, which would make the contours tilt upwards, with a higher slope than the mean density line. The shapes and orientations of these contours reflect the gravitational clustering of both matter and halos.
							
			\subsection{Global one-point statistics}
			\label{ssec:g1p}
			
			Consider a dark matter particle of mass $M_\mathrm{p}$ in an N-body simulation, and let this particle's comoving Voronoi volume be $\V$. We define its Voronoi density to be,
			\begin{align}
				\rho_{\V} = \frac{M_\mathrm{p}}{\V}\, .
			\end{align}
			 Rather than defining density contrasts with respect to the mean Voronoi density, we define,
			 \begin{align}
			 \label{eq:dcm}
			 \delta_{\V} = \frac{\rho_{\V}}{\brho} - 1\, ,
			 \end{align}
			where $\brho$ is the usual mean comoving Eulerian matter density. In a simulation box with sides of length $L_{\mathrm{box}}$ and $N_\mathrm{p}$ particles, $\brho = N_\mathrm{p}\, M_\mathrm{p}\, L_\mathrm{box}^{-3}$. The Voronoi density contrast defined in this way is the fractional difference between the mean volume per particle and the volume of an individual particle's Voronoi cell. At early enough times every particle, to a good approximation, has the same Voronoi volume. Then the Voronoi density contrast can be interpreted as the fractional amount of change in an individual particle's Voronoi volume since early times. 
			
			The mean Voronoi mass density is not equal to the mean mass density, $\bar{\rho}$, but rather,
			\begin{align}
				\bar{\V} = \frac{M_\mathrm{p}}{\bar{\rho}} \, .
			\end{align}
			The Voronoi density contrasts can be expressed in terms of the mean Voronoi volume, $\bar{\V}$, as
			\begin{align}
				\delta_{\V} = \frac{\bar{\V}}{\V} - 1 \, .
			\end{align}			
			The Voronoi density contrasts do not average to zero, but their volume weighted average vanishes,
			\begin{align}
				\label{eq:vwa}
				\big\langle \delta_\V\, \V \big \rangle = 0\, .
			\end{align}
			We denote the probability of finding a cell with Voronoi density between $1 + \delta_\V$ and $1 + \delta_\V + \mathrm{d}\delta_\V$ as
			\begin{align}
				\pdf(1 + \delta_\V)\, \mathrm{d}\delta_\V\, .
			\end{align}
			We refer to $\pdf(1 + \delta_\V)$ as the Voronoi density PDF. We will now consider how the shape of this distribution, measured in a finite subregion of the Universe, differs from its shape measured globally.
			
			\subsection{Local one-point statistics}
			\label{ssec:l1p}
			
			In a finite subregion, the local estimation of the Voronoi density PDF will differ from the globally measured PDF. The spatial variations in the locally measured PDF are due to spatial variations in the local matter density. For a subregion of volume $V_{\su}$ centered at position $\vec{x}$, a local observer sees a mean density, 
			\begin{align}
				\brho_\su(\vec{x}) = \frac{1}{V_\su} \int_{V_\su} \!\! \dee^3 y\, \rho(\vec{y})\, .
			\end{align}
			If the region is large enough, the local mean density will differ from the global one by a perturbatively small fluctuation, $\dl$, 
			\begin{align}
					\brho_\su(\vec{x})  = \brho \lb 1 + \dl(\vec{x}) \rb\, .
			\end{align}
			The subscript ``su'' is meant to indicate ``separate universe''. From the point of view of an observer who lives within and only has observational access to $V_\su$, they ascribe $\brho_\su$ as the mean density of the Universe. This observer would therefore disagree about the value of cosmological parameters with a global observer, who has wider observational access to regions outside of $V_\su$. 
			
			We refer to the observer limited to $V_\su$ as a \emph{separate universe observer}, and all of the observables measured with respect to their coordinates and cosmology are \emph{separate universe observables}. For example, if $\delta_L$ is positive and large enough, then the separate universe observer determines that they live within a closed universe, when in actuality they may only live within a collapsing subregion of an otherwise open universe.
			
			The local expansion history and structure formation occur within the separate universe observer's reference frame as if they live in a Friedmann-Lema\^itre-Robertson-Walker cosmology, but with parameters shifted from their global values in a way that depends on the evolution of $\dl$
			\cite{Sirko_2005, Hu:2016ssz}. This implies that all differences between the separate universe observables and their counterparts in the global cosmology should be characterized by their response to the presence and growth history of $\dl$. We refer to this statement as the \emph{separate universe ansatz}.
			
			Voronoi density contrasts measured by a separate universe observer are defined,
			\begin{align}
				\delta_{\V, \su} =\frac{\bar{\V}_\su}{\V}- 1 \, ,
			\end{align}
			where $\bar{\V}_\su$ is the local mean Voronoi volume within the subregion, including only particles whose positions lie within $V_\su$ \footnote{The total Voronoi volume may not sum to $V_\su$ due to fluctuations at the boundary, depending on how this is defined. However, if the region is large enough, these fluctuations are negligibly small.}. This definition preserves the volume weighted average of Eq.~(\ref{eq:vwa}) within the separate universe cosmology. The locally measured Voronoi PDF is the probability of finding a cell within $V_\su$ that has a density between $1 + \delta_{\V,\su}$ and $1 +  \delta_{\V,\su} + \dee \delta_{\V,\su}$. We denote this,
			\begin{align}
				\pdf_{\su}(1 + \delta_{\V, \su}|\vec{x})\, \dee\delta_{\V,\su} \, .
			\end{align}		
			
			The separate universe ansatz states that the spatial modulations of local observables are due to the effects of the large-scale density fluctuations, $\dl(\vec{x})$, so that,
			\begin{align}
				 \pdf_{\su}(1 + \delta_{\V, \su}\,|\,\vec{x}) = 	\pdf_{\su}(1 + \delta_{\V,\su}\,|\,\dl(\vec{x})) \, .
			\end{align}
			For a large enough subregion, we can expand this around $\dl = 0$, keeping only the leading term,	
			\begin{align}
				\pdf_{\su}(1 + \delta_\V\,|\,\dl) = \pdf(1 + \delta_\V) + \frac{\dee\pdf_\su}{\dee\dl} \dl(\vec{x}) + \mathcal{O}(\dl^2) \, .
			\end{align}
			On the left-hand side of this equation, $\delta_{\V,\su} = \delta_\V$ in the limit $\dl\rightarrow 0$, so we have dropped that subscript. The higher order terms will also include tidal contributions $\sim \mathcal{O}(\nabla^2\dl)$, which enter at second order. The fractional difference between the local and global PDF gives its linear response to $\dl$, which can be measured by correlating the fluctuations of $\pdf_\su$ with the long-wavelength density perturbations,
			\begin{align}
				\label{eq:rc}
				\frac{\dee \log \pdf_\su}{\dee \dl} = \frac{\big \langle\log \big( \pdf_\su(\vec{x})\big) \, \dl(\vec{x}') \big\rangle}{\big \langle \dl(\vec{x})\, \dl(\vec{x}') \big \rangle} \, .
			\end{align} 
			We refer to the quantity as the separate universe response of the Voronoi density PDF.
			
			\subsection{Halos}
			\label{ssec:vph}
			
			The strength of the correlation between a halo's virial mass and the total mass within its Voronoi cell depends strongly on the halo's virial mass. For low-mass halos, the two masses are very weakly correlated, while for high-mass halos the correlation is strong. Observationally, the amount of mass within a Voronoi cell is not easily known. However, it may be related to the number density of halos through a nonlinear bias model. For these reasons, we measure the halo number density PDF, rather than a halo mass density PDF.
			
			For a single Voronoi cell of volume $\V$, the number density is,
			\begin{align}
				n_{\V} = \frac{1}{\V} \, ,
			\end{align}
			and we define the Voronoi density contrast,
			\begin{align}
				\label{eq:dch}
				\delta_{\V} = \frac{n_{\V}}{\bar{n}} - 1 \, ,
			\end{align}
			where $\bar{n}$ is the mean Eulerian halo number density. The probability of finding a halo Voronoi cell with density between $1 + \delta_\V$ and $1 + \delta_\V + \mathrm{d}\delta_\V$ is denote $\pdf(1 + \delta_\V)\, \mathrm{d}\delta_\V$. Note that this notation is the same as for the matter Voronoi PDF, but we will always specify which one we are dealing with to avoid confusion. 
			
			In a large, finite subregion of volume $V_\su$ centered at $\vec{x}$, the local mean number density differs from the global mean,
			\begin{align}
				\bar{n}_\su(M, \vec{x}) & = \bar{n}(M)\big( 1 + \delta_n(M,\vec{x})\big)  \nonumber \\
				& \simeq \bar{n}(M)\big( 1 + b_h(M)\,\dl(\vec{x}) \big) \, ,
			\end{align}
			where, as before, $\dl$ is the large-scale matter fluctuation within $V_\su$, $b_h$ is the linear Eulerian halo bias, and $\delta_n$ is the local fluctuation in halo number density. In terms of the local mean number density, the Voronoi density contrasts are defined,
			\begin{align}
				\delta_{\V,\su} = \frac{n_\V}{	\bar{n}_\su} - 1 \, .
			\end{align}
			We use the same notation for the position-dependent Voronoi PDF and its separate universe response for both halos and dark matter.
			
			Suppose we measured the position-dependent PDFs with respect to the global mean density (i.e. the density contrasts defined in Eqs.~(\ref{eq:dcm}) and (\ref{eq:dch}), and denote these local PDFs as $\pdf_\mathrm{loc}$. The transformation from $\pdf_\su$ is given by
			\begin{align}
				\pdf_{\mathrm{loc}} = 	\pdf_{\su}\lb 1 - b_h\, \dl \lb 1 + \frac{\dee \log \pdf}{\dee \log (1 + \delta_\V)} \rb \rb \, ,
			\end{align}
			and the linear responses to $\dl$ are related,
			\begin{align}
				\label{eq:loc}
				\frac{\dee\log\pdf_{\mathrm{loc}}}{\dee \dl} = \frac{\dee\log\pdf_\su}{\dee \dl} - b_h \lb 1 + \frac{\dee \log \pdf}{\dee \log (1 + \delta_\V)} \rb \, .
			\end{align}
			The second term on the right-hand side can be attributed to the shift in the local mean background density relative to the global mean density. It does not depend on the growth history of $\dl$ and can be determined directly from the global PDF. The same expression holds for the transformation of the matter PDF, setting the bias equal to one.
			
			\begin{table}
				\begin{tabular}{c c}
					\hline\hline
					Parameter & Value \\
					\hline
					$\oL$ & 0.7 \\			
					$\om$ & 0.3 \\
					$\ob$ & 0.05 \\
					$h$ & 0.7 \\
					$n_\mathrm{s}$ & 0.968 \\
					$A_\mathrm{s}$ & 2.137$\times10^{-9}$ \\ 
					$N_\mathrm{p}$ & $(1024)^3$ \\
					$L_\mathrm{box}$ & $1\ \rm{Gpc}/$$h$ \\
					$M_\mathrm{p}$ & $1.108\times10^{11}\ \msun$\\
					\hline
					\hline
				\end{tabular}
				\caption{Cosmological and N-body simulation parameters.}
				\label{tab:cos}
			\end{table}
			
		\section{Simulations}
		\label{sec:sim}
	
			We ran three sets of 20 N-body simulations, for a total of 60. One set was run with the global expansion history from a standard $\lcdm$ cosmology. The second set, corresponding to an overdense separate universe, was given a slightly modified expansion history, according to Eqs.~(\ref{eq:t1})--(\ref{eq:t2}). The long-wavelength mode $\dl$ evolved linearly as a growing, adiabatic perturbation normalized to $\dl=+0.01$ at the final redshift $z=0.00$. The third set, corresponding to an underdense separate universe, had $\dl=-0.01$ at the final redshift.
			
			All three sets used the same 20 random seeds to generate their initial conditions, which ensures cosmic variance largely cancels at leading order when computing separate universe response observables. The initial conditions were generated using the second order Lagrangian perturbation theory method \cite{Crocce:2006ve}, with input power spectrum from CLASS \cite{Blas:2011rf} using the parameters in Table \ref{tab:cos}. Our simulations were run using a modified version of Gadget2 \cite{Springel:2005mi}, which uses tabulated values of the local separate universe scale factor and Hubble rate, rather than integrating the Friedmann equation to compute the expansion history. Our simulation box size was chosen to be $L_{\mathrm{box}} = 1\ \mathrm{Gpc}/h$ with $N_\mathrm{p} = (1024)^3$ dark matter particles. The particle mass was $M_\mathrm{p} = 1.108\times 10^{11}\ \msun$. These simulations were previously used for a study of void bias in the separate universe \cite{Jamieson:2019dmp}, and the position-dependent Eulerian density PDF \cite{Jamieson:2020wxf}. 
			
			Our halo catalogs were extracted using the friends-of-friends halo finder Rockstar \cite{Behroozi:2011ju}, modified so that the density thresholds used for spherical overdensity mass calculations were consistent with the different separate universe cosmologies, and so that the mass assignments were continuous rather than discrete (see \cite{Li_2016, Jamieson_2018} for details). We measured the Voronoi number density field in a catalog of halos with lower mass cutoff $M > 10 ^{13}\ \msun$, corresponding to a minimum of 100 particles per halo.
		
			Our Voronoi tessellations were carried with a C++ code that uses the Voro++ publicly available library \cite{doi:10.1063/1.3215722}. The object-oriented Voro++ library stores particles in container objects, which we defined to be subboxes of our simulations. We included a buffer region in our subbox containers, keeping track of which particles were in the subbox and which were in the buffer. The Voro++ library includes a loop class, which enabled us to easily compute only the Voronoi volumes of particles in the subbox, including ones with neighbors that are in the buffer. If the buffer is taken to be too small, a particle in the subbox may end up with one of the container walls as its neighbor. In this case, the code throws and error and exits, indicating that the size of the buffer region needs to be increased. 
			
			The advantage of our code is that it is easily parallelizable, since each subbox container can be analyzed independently. It is also has efficient memory usage, since only particles in the containers that are currently being analyzed are loaded into memory. This allowed us to compute the Voronoi volumes of all $1024^3$ dark matter particles in our simulations in 25 minutes, running on 3 compute nodes with 24 cores per node. We tested our code by comparing to the tessellation from ZOBOV \cite{Neyrinck:2007gy}, and found that the Voronoi volumes agree within floating point accuracy. Our code uses MPI to distribute processes to each node, and OMP to distribute processes within a node to each core.
		
			\subsection*{Kernel density estimator}
			\label{ssec:kde}
			
				The Voronoi tessallations around our simulation dark matter particles and halos assign a volume, which we converted to a density, for each point-like tracer. From this list of Voronoi densities, we computed the PDFs using a kernel density estimator (KDE). First, we converted the lists of Voronoi densities into lists $\big\{\!\log(1 + \delta_{\V, i})\big\}_{i=1}^{N}$, and used a Gaussian kernel,
				\begin{align}
				\pdf_{\log}\big( \!\log ( & 1 + \delta_\V)\big) = \nonumber \\ & \frac{1}{N w} \sum_{i=1}^{N} \exp\lb-\frac{1}{2 w^2} \log\lb\frac{1+\delta_\V}{1+\delta_{\V,i}}\rb^2 \rb \, .
				\end{align}
				Here, $w$ is the kernel width, which sets the level of smoothing over each data point, and $\pdf_{\log}$ is the PDF for $\log(1 + \delta_{\V})$. We then transformed this to $\pdf(1 + \delta_{\V})$,
				\begin{align}
				\label{eq:plog}
				\pdf(1+\delta_\V) = \frac{\pdf_{\log}\big( \!\log (1 + \delta_\V)\big)}{1+\delta_\V} \, .
				\end{align}
				This procedure is justified by the observation that the density PDFs are, to a first approximation, similar to $\log$-normal distributions \cite{Bernardeau:1994aq, Coles:1991if}.
				
				The choice of kernel width is import (the kernel function itself is less important), because it sets the level of smoothing, which controls the variance and statistical bias of the estimator. In the asymptotic limit of large sample size, the variance and estimator bias for a Gaussian KDE are \cite{Silverman_1986},
				\begin{align}
				\mathrm{Var}(\pdf_{\log})^2 \simeq \frac{1}{N w}\pdf_{\log}\, , \\
				\label{eq:kdeb}
				\mathrm{Bias}(\pdf_{\log}) \simeq \frac{w^2}{2}\frac{\dee^2 \pdf_{\log}}{\dee \log(1 + \delta_\V)^2} \, .
				\end{align}
				If the PDF $\pdf_{\log}$ were perfectly Gaussian, the optimal width given by minimizing the integral of the squared asymptotic error would be,
				\begin{align}
				w_{\mathrm{G}} = \sigma_{\pdf_{\log}} \lb \frac{4}{3 N} \rb^{1/5} \, ,
				\end{align}
				with $ \sigma_{\pdf_{\log}}$ being the standard deviation of $\pdf_{\log}$, which can be estimated from the sample. For the dark matter, the PDF in the global cosmology prefers a broader width than a Gaussian distribution, so we chose kernel widths $w = 1.5\, w_{\mathrm{G}}$. For the halos, the global PDF prefers a narrow width of $w = 0.3\, w_{\mathrm{G}}$, as we demonstrate in the \hyperref[sec:app]{Appendix} to this paper, where we discuss these width selections in greater detail.
				
				Estimating derivatives of PDFs using a KDE permits much more smoothing. Since taking a derivative essentially involves subtracting nearly identical functions, the estimator bias largely cancels in the numerator. This is true for the linear responses of the PDF to $\delta_L$, which are approximated in the separate universe formalism as finite difference derivatives. Minimizing the integral of the asymptotic squared error for the derivative of a Gaussian PDF, the optimal width scales as $N^{1/7}$ rather than $N^{1/5}$. We chose widths of $w_\su = 2.5\, w_{\mathrm{G}}$ for the separate universe PDFs of the dark matter, and $w_\su = 4.5\, w_{\mathrm{G}}$ for the halos. We present tests of the effects from varying the separate universe smoothing for the halos in the \hyperref[sec:app]{Appendix}. For the dark matter, the effects of varying the smoothing width are smaller, due to the large sample size.
				
				\begin{table}
					\begin{tabular}{|c||c|c||c|c|c||c|}
						\hline
						\multicolumn{7}{|c|}{Dark matter} \\ \hline
						$z$ & $\mu$ & $\sigma$ &  $q_1$ & $q_2$ & $q_3$ & $\delta_\V>0$ \\ \hline
						0.0 & 1170 & 7123 &  0.834 & 5.63 & 132 & 0.722 \\
						0.5 & 382.9 & 2565 & 0.710 & 2.79 & 34.3 & 0.681 \\
						1.0 & 157.4 & 1185 & 0.665 & 1.85 & 10.9 & 0.642 \\
						49 & 1.004 & 0.062 & 0.960 & 1.00 & 1.04 & 0.505 \\  \hline
					\end{tabular}
				
					\vspace{0.03in}
					
					\begin{tabular}{|c||c|c||c|c|c||c|}			
						\hline
						\multicolumn{7}{|c|}{Halos, $M > 10^{13}\ \mathrm{M}_\odot$ } \\ \hline	
						$z$ & $\mu$ & $\sigma$ &  $q_1$ & $q_2$ & $q_3$ & $\delta_\V>0$ \\ \hline		
						0.0 & 2.403 & 3.292 & 0.760 & 1.42 & 2.77 & 0.644 \\
						0.5 & 2.467 & 3.559 & 0.750 & 1.40 & 2.78 & 0.638 \\
						1.0 & 2.610 & 4.316 & 0.739 & 1.36 & 2.78 & 0.629 \\ \hline \hline
					\end{tabular}				
					\caption{Properties of the Voronoi density PDFs. The left column lists the redshift, followed by the distribution mean $\mu$ and standard deviation $\sigma$. The next three columns list the first, second and third quartiles. That is, $q_i$ is the value of $1 + \delta_\V$ for which $i \times 25\%$ of the cells have a lower density. The final column gives the fraction of cells that are overdense.}
					\label{tab:dmdist}
				\end{table}
						
			\begin{figure*}
				\graphic{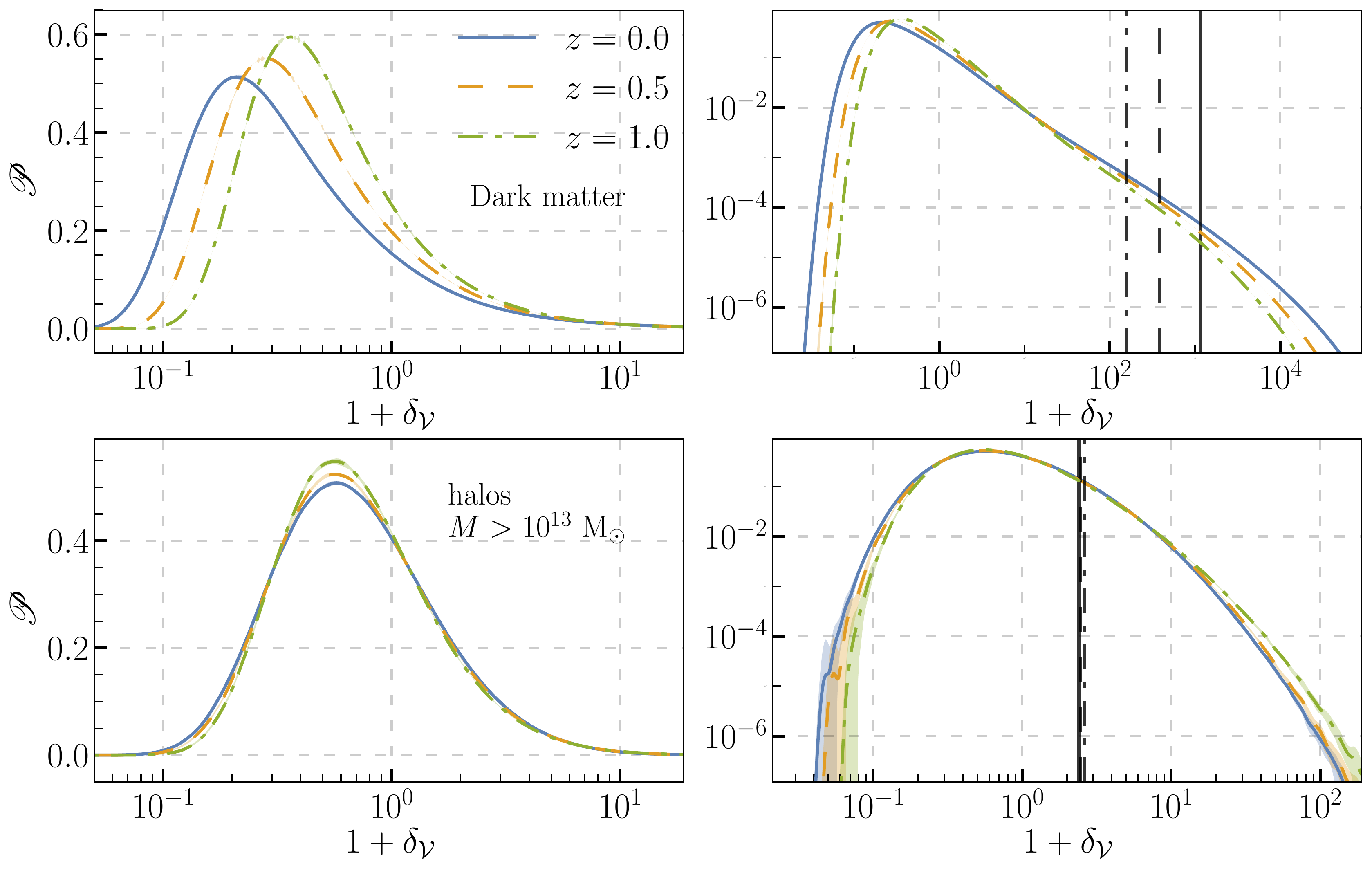}{0.98}
				\caption{Top: Dark matter density probability distributions measured with respect to Voronoi tessellations around N-body simulation dark matter particles. The plots on the left have a linear vertical scale to demonstrate the evolution of the distribution near the peak, while the plots on the right have a logarithmic vertical scale to demonstrate the evolution of the tails of the distributions.  Bottom: Halo number density probability distributions measured with respect to Voronoi tessellations around halos with masses $M> 10^{13}\ \mathrm{M}_\odot$. The legend on the top left plot, indicating the redshift of each distribution, applies to all plots. The black vertical lines indicate the mean of the distribution, with dashes that correspond to redshift according to the legend. The $1\sigma$ bootstrap errors are plotted as shaded regions, but are too small to see except in the extreme tails.}
				\label{fig:pdfs}
			\end{figure*}

		\section{The Global Voronoi density PDF}
		\label{sec:gvp}
		
				The mean volume per dark matter particle is constant in N-body simulations. For our simulations, the value is fixed at $\bar{\V} = 2.72\ \mathrm{Mpc}^3$. At early times, the distribution of matter cell densities is nearly Gaussian, narrowly peaked around $\rho = \bar{\rho} \equiv M_\mathrm{p}/\bar{\V}$. The exact distribution cannot be Gaussian, since $\rho$ is bounded from below, however the width of the distribution is sufficiently narrow so that the non-Gaussianity is small. In fact, since we began our simulations at redshift $z=49$ and used second order Lagrangian perturbation theory to set up the initial conditions, the initial density PDF is better approximated as log-normal. The interplay of cosmic expansion and gravitational collapse concentrate most of the dark matter into overdense regions, producing a long, high-density tail in the PDF at late times. The dark matter Voronoi PDFs measured for particles with mass $M_\mathrm{p} = 1.1\times 10^{11}\ \msun$ are show in Fig.~\ref{fig:pdfs} at redshifts $z=0.0,~0.5,$ and 1.0. The location of the peak decreases with redshift, and occurs at very low densities near redshift $z=0.0$. While the distribution peaks at $\delta_\V < 0$, the long tail at high densities ensures that a majority of the particles are in overdense cells, with   $\delta_\V  > 0$. The vertical lines in Fig.~\ref{fig:pdfs} indicate the mean of each distribution.
		
				Some properties of the dark matter and halo PDFs are summarized in Table \ref{tab:dmdist}. The mean of the dark matter PDF increases by nearly an order of magnitude between redshifts $z=1$ and $z=0$. The mean Voronoi density is $77\%$ of the virial density for spherical collapse at redshift $z=1$ (148 $\bar{\rho}$) , while the mean is 3.5 times larger than the virial density at $z=0$ (337 $\bar{\rho}$). Many of the dark matter particles within the virial radius of a halo have cell densities that are lower than the virial density, and while the individual Voronoi cells are typically far from spherical, we will demonstrate in Sec. \ref{sec:mod} that the virial density for spherical collapse still appears as an important scale in the dark matter Voronoi PDF.
				
				The number of halos is not constant over time, so the mean Voronoi volume per halo also evolves with redshift. The halo Voronoi density PDF captures information about the local environment of halos and their clustering with each other. Unlike the spatial correlation function, which is computed using the distances of each halo from all other halos in the catalog, the Voronoi PDF characterizes only the nearest neighbors of each halo. The halo Voronoi density PDFs are shown in Fig.~\ref{fig:pdfs} for halos with $M > 10^{13}\ \msun$. The peak of the distribution evolves considerably less than that of the dark matter distribution. The clustering of halos is made evident by the broadening of the distribution with time, as more collapsed objects are concentrated, with smaller cells, into high-density regions, leaving halos with larger Voronoi cells in lower density regions. However, the clustering also causes mergers, which combine high-density cells into single, somewhat larger cells for high-mass halos. This can be seen from the decrease in the high-density tail of the halo PDF with time.
					
		\section{The Local Voronoi density PDF}
		\label{sec:lvp}
		
			\begin{figure*}
				\graphic{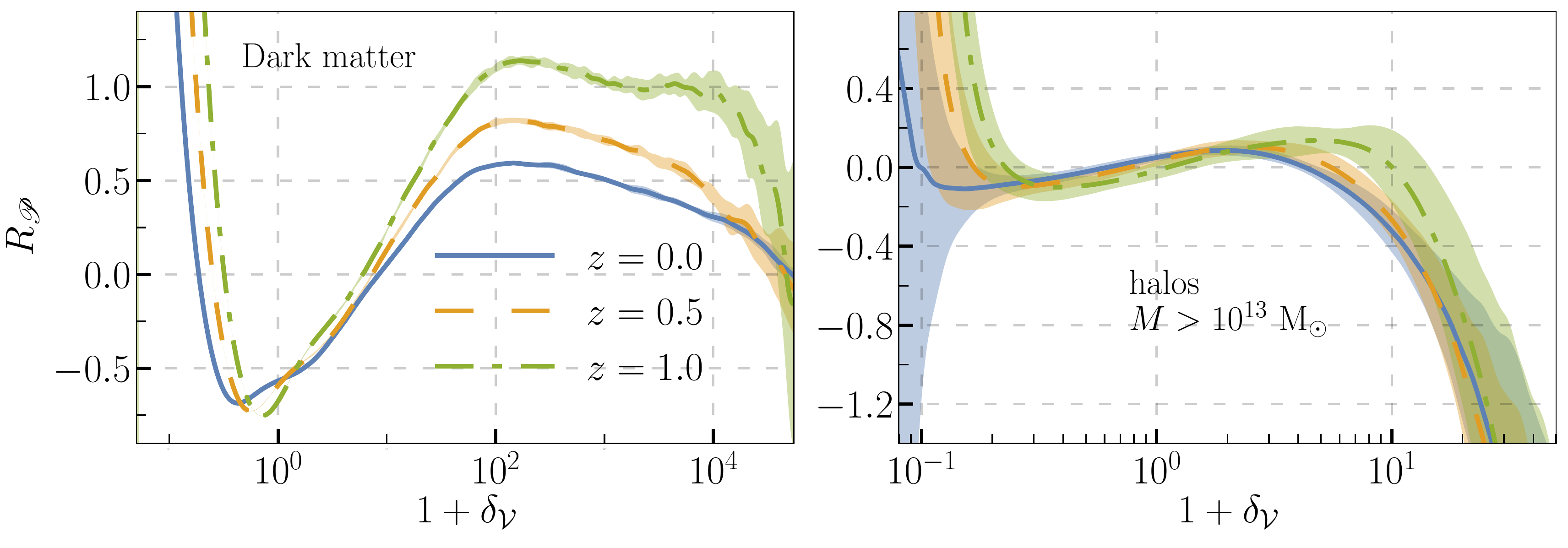}{1.}
				\caption{Separate universe responses of the Voronoi density PDF of dark matter (left) and halos (right), at redshifts $z=0.0$, 0.5, and 1.0, measured in N-body simulations. The smoothing scale for the dark matter is given by the particle mass $M_\mathrm{p}=1.1\times10^{11}~\msun$. For the halos, a lower mass cutoff of $10^{13}~\msun$ was used to construct the catalogs. The shaded regions correspond to $1\sigma$ bootstrap errors.}
				\label{fig:Rsu}
			\end{figure*}
		
			\subsection{Separate universe formalism}
			\label{ssec:suf}
			
				The dynamics of gravitational clustering is nonlinear, so the long-wavelength perturbations of the density field couple to, and effect the dynamics of structure formation on smaller scales. In the separate universe formalism, these effects are computed by defining a local cosmology in which the long-wavelength density perturbations are absorbed into the background, homogeneous density,
				\begin{align}
					\bar{\rho}_{\su} = \bar{\rho}\lb 1 + \dl \rb \, .
				\end{align}
				The locally defined mean density, $\brho_\su$, yields a locally defined cosmology that differs from the global one due to the presence of $\dl$, the large-scale fluctuations. The local scale factor and local Hubble rate can be computed to linear order in $\dl$ \cite{Sirko_2005} as
				\begin{align}
					\label{eq:t1}
					a_\su & \simeq a \lb 1- \frac{1}{3} \dl \rb  \, , \\
					\label{eq:t2}
					H_\su & \simeq H \lb 1 - \frac{1}{3} \dl' \rb  \, ,
				\end{align}	
				where $'$ indicates $\dee/\dee\log a$.
				
				Since $\dl$ includes only linear modes, its evolution can be computed in the global cosmology, using perturbation theory. From the above equations, we can define a pair of cosmologies, one corresponding to an overdense region ($\dl>0$), and one corresponding to an underdense region ($\dl<0$). By running pairs of N-body simulations with the corresponding overdense and underdense expansion histories, the linear response of an observables in the separate universe $\mathcal{O}_{\su}(\dl)$ can be determined by taking the finite difference derivative with respect to $\dl$,
				\begin{align}
					R_{\mathcal{O}} & = \frac{\dee\log\mathcal{O}_\su}{\dee\dl}\bigg|_{\dl=0} \\
					& \simeq \frac{\mathcal{O}_\su(+\dl) - \mathcal{O}_\su(-\dl) }{2\dl\mathcal{O}} \, ,
				\end{align}
				where $\mathcal{O}$ is the observable in the global universe. If we use the same set of initial conditions in each separate universe pair, we also benefit from cosmic variance cancellation in determinations of $R_\mathcal{O}$. 
				
				The separate universe response of the halo mass function is the linear halo bias \cite{Li_2016, Baldauf_2016}. Similarly, the linear void bias is given by the separate universe response of the void size function \cite{Chan:2019yzq, Jamieson:2019dmp}. Higher order terms in the bias expansion can also be measured using separate universe techniques. A generalized bias can be associated to any observable, $\mathcal{O}$, that is affected by long-wavelength density perturbations. In a previous paper, we studied the generalized bias associated with the Eulerian matter density PDF using the separate universe method \cite{Jamieson:2020wxf}. In the following subsections we present the generalized bias of the Voronoi density PDF in dark matter and in halos.
				
			\subsection{The Voronoi density PDF in the separate universe}	
			\label{ssec:sur}
			
				The separate universe response of the Voronoi density PDF is defined as
				\begin{align}
					\label{eq:rsu}
					R_\pdf = \frac{\pdf_\su(+\dl) - \pdf_\su(-\dl)}{2 \pdf \dl}\, ,
				\end{align}
				where $\pdf_\su(\pm\dl)$ is the local Voronoi density PDF measured in an overdense or underdense separate universe cosmology. Measurements of $R_\pdf$ from our simulations are shown in Fig.~\ref{fig:Rsu} for both the dark matter particles and the halos.
				
				The matter PDF response has an intuitive density dependence. The response is positive for low-density cells ($\delta_\V < -0.5$) and high-density cells ($\delta_\V > 10$), indicating that the tails of the PDF are enhanced in an overdense region, and reduced in an underdense region. The response is negative in cells near the background density ($-0.5 < \delta_\V < 10$), indicating the distribution's peak is enhanced in an underdense region, and reduced in an overdense region. This density dependence reflects gravitational clustering, which concentrates most of the matter into high-density cells, leaving behind a few isolated, low-density cells. Clearly, the response is dynamical in nature, depending on the details and evolution of $\dl$, which is sensitive to cosmology. 
				
				The valley of the PDF response near its minimum is narrower at earlier redshifts, although it is not significantly deeper at $z=1.0$ than it is at $z=0.0$. On the other hand, the positive responses are stronger at earlier times. This reflects the relative rarity of extreme density regions at early times compared to later times. This is analogous to the increase in halo bias at fixed mass with increasing redshift.
				
				Halos have a similar PDF response shape, although with less pronounced features. The response is positive for overdense cells with $0<\delta_\V< 10$  and for underdense cells with $\delta_\V<-0.8$ at redshift $z=1$. These bounds evolve toward lower densities with decreasing redshift. The negative response regime spans a more underdense range compared with the matter response, especially at lower redshift. This is due to the lack of evolution in the position of the halo distribution's peak, which can be seen in Fig.~\ref{fig:pdfs}. An overdense separate universe has enhanced clustering, and can be thought of as being dynamically more advanced in time than the mean universe. For the matter PDF, the peak migrates deeper into the underdense region with time as the overall distribution broadens. Both of these effects contribute to the negative response near $\delta_\V\simeq 0$ for the matter. For the halos, the peak does not significantly change position over time, and so the response is negative at lower densities, although this depends on the halo mass threshold.
				
				For the halos, the response at high Voronoi density ($\delta_\V > 10$) becomes negative, indicating that there are more extremely high-density cells in an underdense compared to an overdense separate universe. This may seem counter intuitive, but it is simply due to the merger of halos. Extremely high-density neighboring cells coalesce and becomes one larger cell as halos merge. Since the overdense cosmology can be thought of as advanced in time with respect to clustering, more of these mergers have occurred at a fixed global time compared with the underdense separate universe, so the response is negative. This interpretation is consistent with the redshift evolution of the global PDF in Fig.~\ref{fig:pdfs}, which shows the high-density tail decreasing with time.
				
				Since both $\pdf$ and $\pdf_\su$ are normalized, this implies,
				\begin{align}
					\label{eq:IRP}
					\int_{-1}^{\infty} \! \dee \delta_\V \, R_\pdf \pdf = 0 \, .
				\end{align}
				The orthogonality of the PDF and its linear response allows us to further interpret the physical meaning of the PDF response. To see this, we first note that we can recast the halo Voronoi density PDF as a conditional mass function, 
				\begin{align}
					n(M, 1+\delta_\V) =  \bar{n}(M) \pdf(1 + \delta_\V) \, ,
				\end{align}
				where $\bar{n}(M)$ is the cumulative halo mass function. Note that $\pdf$ also depends on the lower mass threshold since, if we vary the threshold, we change the halo catalog and must redo the tessellation. The linear response of this conditional halo mass function is,
				\begin{align}
					\frac{\dee\log n_\su(M,1+\delta_\V)}{\dee\dl} = b_h(M) + R_\pdf(1 + \delta_\V)\, .
				\end{align}
				The left-hand side of this expression is the generalized bias of the conditional halo mass function, for halos selected based both on their Voronoi density and a lower mass threshold. Halos of the same mass but different Voronoi cell size have different biases, so this generalized bias can be interpreted as a particular kind of assembly bias. From Eq.~(\ref{eq:IRP}), we find
				\begin{align}
					\frac{1}{\bar{n}(M)} \int_{-1}^{\infty}\!\dee \delta_\V \, \frac{\dee \log n_\su(M, 1 + \delta_\V)}{\dee\dl} n(M,1+\delta_\V) = b_h \, .
				\end{align}
				The generalized bias of the conditional halo mass function decomposes the total halo bias into contributions from halos of different Voronoi density. The interpretation of the dark matter PDF response is the same, except in that case the bias is unity, and the dark matter PDF response gives the bias of equal-mass regions of the density field selected based on their Voronoi density.
	
				\begin{figure*}
					\centering
					\graphic{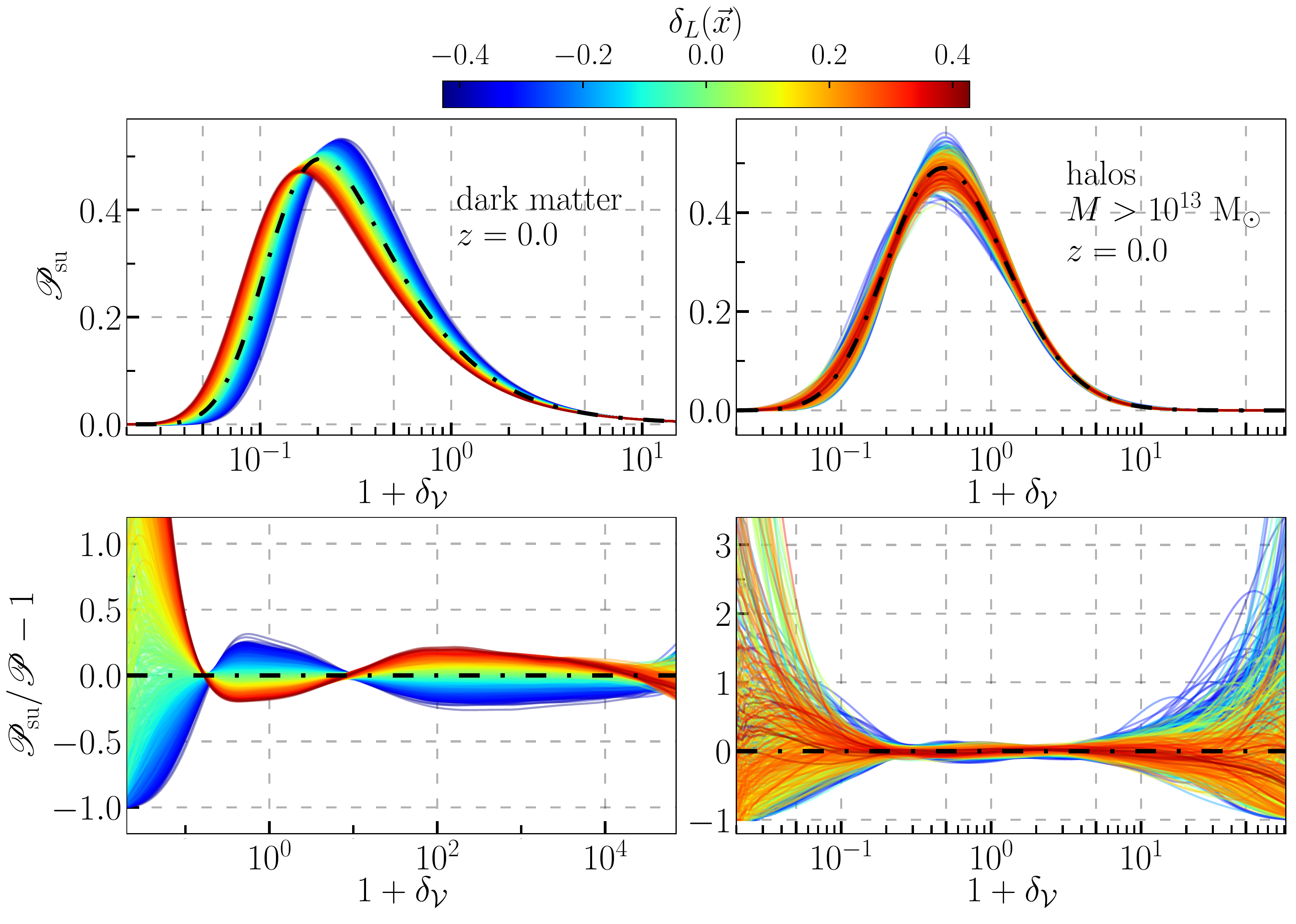}{1.0}
					\caption{Local Voronoi density PDFs for the matter (left) and halos (right) at redshift $z=0.0$, measured in subboxes with sides lengths of $125~\mpc/h$ within a global universe simulation. The color of each curve corresponds to the value of the large-scale density fluctuation within each subbox.}
					\label{fig:lpdfs}
				\end{figure*}
				
				\begin{figure*}
					\centering
					\graphic{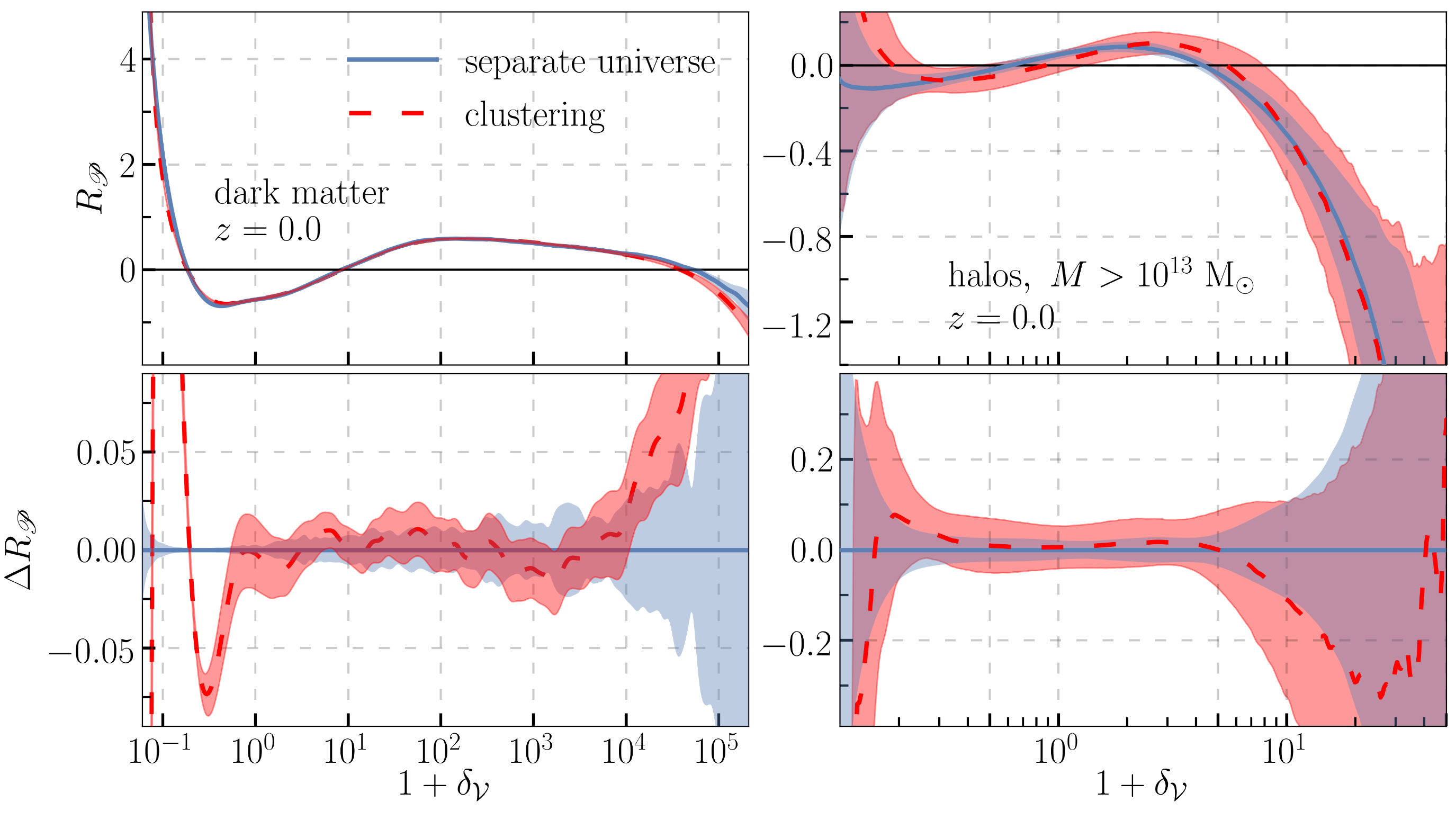}{1.0}
					\caption{Comparison between the separate universe response and the clustering bias of the Voronoi density PDFs for matter (left) and halos (right) at redshift $z=0.0$. The difference between the separate universe responses and clustering biases are shown in the bottom panels.}
					\label{fig:suvc}
				\end{figure*}
		
			\subsection{Validation in the global universe}
			\label{ssec:guv}
			
			The separate universe ansatz claims that the spatial modulations in local, position-dependent observables measured in finite subregions are due to the effects of long-wavelength modes on gravitational clustering. To test this, we measured the local Voronoi density PDF in subboxes of global universe simulations, and cross-correlated the fluctuations of the local PDF with the matter field.
			
			First, we constructed a mesh of $32^3$ overlapping cubic subboxes with sides of length $125~\mathrm{Mpc}/h$. In each subbox, we load the N-body particles from one of our global universe simulations with positions that lie within the subbox, along with their Voronoi densities. We computed the local PDF $\pdf_\su(1 + \delta_{\V,\su}\, |\, \vec{x})$, and the local matter fluctuation, $\delta_L(\vec{x})$, where $\vec{x}$ is the position of the center of the subbox. Next, we computed the local PDF fluctuation,
			\begin{align}
				\frac{\pdf_\su(\vec{x})}{\pdf} - 1 \simeq 
				\log\big(\pdf_\su(\vec{x})/\pdf\big)\, .
			\end{align}
			Here we have suppressed the Voronoi density argument of the PDF. In practice we computed the fractional difference on the left-hand side of the above expression, but we use the logarithmic notation on the right-hand side here for brevity. They are equivalent to leading order in $\dl$.
			
			We used FFTW3 \cite{FFTW05} to compute the matter modes, $\dl(\vec{k})$, and the modes of the PDF fluctuations, $\log\big(\pdf_\su(\vec{k})\big)$.  According the separate universe ansatz, and from Eqs.~(\ref{eq:rc}) and (\ref{eq:rsu}),
			\begin{align}
				R_\pdf = \lim_{k\rightarrow 0} \frac{\big \langle\log \big( \pdf_\su(\vec{k})\big) \, \dl(\vec{k}') \big\rangle}{\big \langle \dl(\vec{k})\, \dl(\vec{k}') \big \rangle} \, .
			\end{align}
			The left-hand side is the separate universe response of the PDF, and we refer to the right-hand side as the clustering bias of the local Voronoi density PDF. To extract this from our simulation data, we first compute $R_\pdf(k)$, which is the clustering bias in the above expression prior to taking the $k\rightarrow 0$ limit. We then fit a linear polynomial in $k^2$,
			\begin{align}
				R_\pdf(k) = b_1 + b_2 k^2\, .
			\end{align}
			and take the constant term $b_1$ as the linear clustering bias. 
			
			A sample of local Voronoi PDFs is shown in Fig.~\ref{fig:lpdfs}, with colors corresponding to the local large-scale density fluctuation. The underdense matter PDFs are clearly narrower, more sharply peaked, and with peak positions nearer to $\delta_\V=0$. The PDFs in overdense subboxes are broader and have their peaks at lower densities. The shapes of the fractional differences between the local and global PDFs, shown in the bottom panel, reproduce the shape of the separate universe response. This is also true for the halos, although this is not as easy to see since the response is weaker in this case, and the local PDFs are noisier due to the lower sample size.
			
			The comparison between the PDF clustering bias and the separate universe PDF response is shown in Fig.~\ref{fig:suvc}. For the matter PDF, the agreement is excellent for $\delta_\V > -0.5$, until around $\delta_\V \simeq 2\times 10^{4}$, spanning nearly five orders of magnitude of Voronoi cell densities. This clearly demonstrates that the separate universe ansatz is quite accurate over an enormous range of densities. Interestingly, the PDF response provides a test that distinguishes between overdense and underdense cells, unlike previous work that has tested the separate universe ansatz using the power spectrum \cite{Li_2014a, Chiang_2014}. Since the power spectrum is quadratic, overdensities and underdensities are treated as equivalent. Our results for the matter Voronoi density PDF response represent a validation of the separate universe method across the widest range of densities to date.
			
			The disagreement at extremely high densities is unsurprising. On the one hand, these densities are so far out into the PDF's tail that their abundances are considerably low. Given the difference in sample sizes between the separate universe simulations and the global universe subboxes, the corresponding PDF tails must incur different statistical biases from the KDE. On the other hand, particles with the smallest Voronoi cells are subject to force softening. We set the softening length to 20~$\mathrm{kpc}$ in our N-body simulations. A spherical cell with  radius equal to this softening length has a density corresponding to $1+\delta_\V = 2.8\times 10^4$. Note, we have not consistently adjusted the comoving softening length in the separate universe simulations to match the physical softening length in the global universe, so it is unsurprising to find discrepancies at these extreme densities.
			
			For underdense cells with $\delta_\V < -0.5$, the clustering bias and separate universe response disagree. This is also not surprising, since the low-density tail of the PDF drops quickly, with the very steep slope, and there is a comparatively smaller sample size of extremely underdense cells. Although the statistical bias incurred from the KDE smoothing on $\pdf_{\log} \big ( \log ( 1 + \delta_\V ) \big)$ is proportional to its curvature (or second derivative), transforming to $\pdf( 1 + \delta_\V )$ yields a contribution to the statistical bias that is proportional to the PDF's slope. We have optimized the KDE widths based on the integrated squared error, which must be suboptimal for the steep, low-density part of the PDF. It may be possible to improve on the KDE using a more sophisticated smoothing, such as choosing local widths as a function of $1+\delta_\V$, but we leave this for future work.
			
			In Fig.~\ref{fig:suvc}, we have also shown the comparison between the halo PDF clustering bias and the separate universe response. The agreement is good across the entire range of densities, spanning three orders of magnitude. The variance of the clustering bias is larger than that of the separate universe response. This is consistent with previous results for halo bias, void bias, and the Eulerian PDF response \cite{Li_2016, Jamieson:2019dmp, Jamieson:2020wxf}. The separate universe method also yields a lower variance for the matter PDF response at densities where $\delta_\V < 10^2$. At higher densities, the clustering method and separate universe method yield similar variances.
			
		\section{Outlook on modeling}
		\label{sec:mod}
		
			\begin{figure*}
				\centering
				\graphic{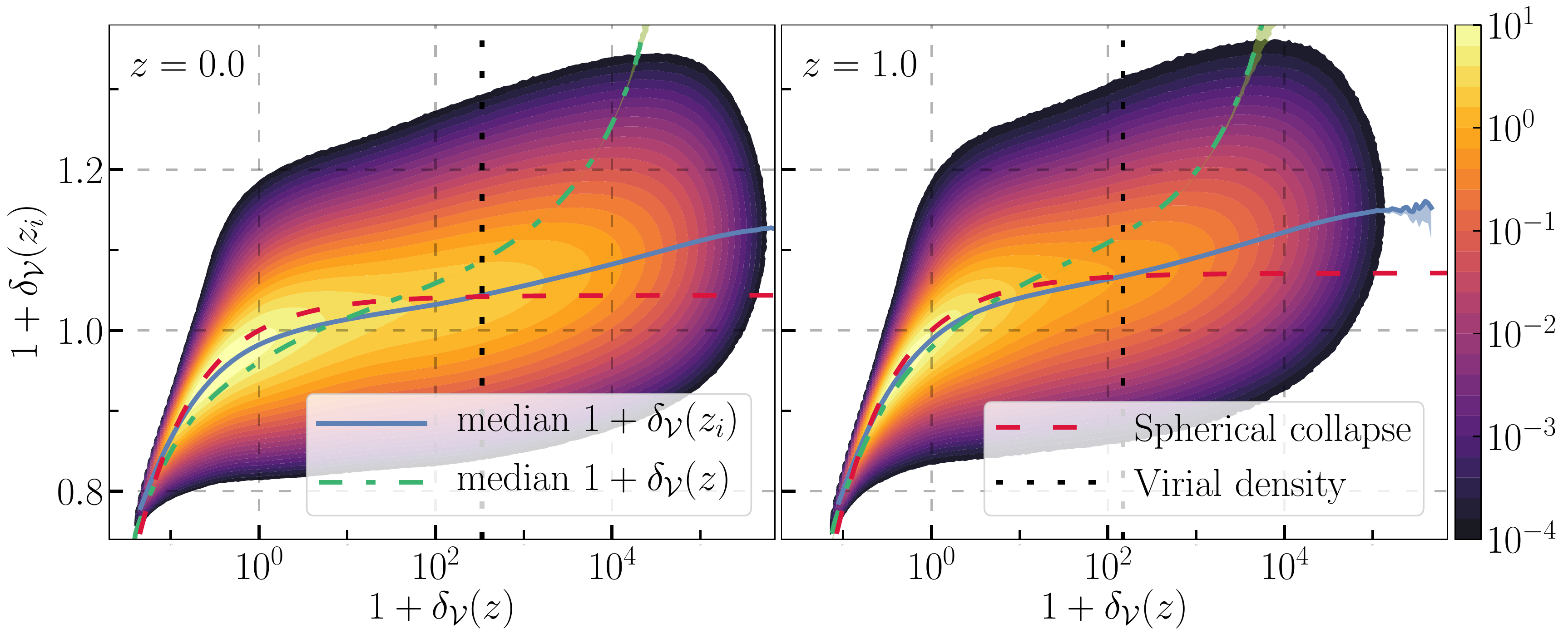}{1.}
				\caption{2D contour plots of the initial and final Voronoi densities of the dark matter particles at final redshifts $z=0.0$ (left) and $z=1.0$ (right). The contours and color bar scale correspond to the normalized bivariate PDF, $\pdf\big(1+\delta_\V(z_i), 1+\delta_\V(z_f)\big)$, estimated from a 2D histogram with 200 uniformly spaced bins along the vertical axis, and 200 logarithmically spaced bins along the horizontal axis. The solid blue curves indicate the median initial density for a fixed final density, while the dash-dotted, light green curves indicate the median final density at fixed initial density. The red dashed lines are the EdS spherical collapse map. The dotted black vertical lines corresponds to the virial density.}
				\label{fig:icf}
			\end{figure*}
		
			In a previous paper, we demonstrated how the Eulerian PDF response of the smoothed matter field could be modeled accurately using only the linear power spectrum and simple, spherical collapse calculations \cite{Jamieson:2020wxf}. For the Voronoi density PDF responses, the situation is much more difficult. For the halos, a typical approach might be to invoke the halo model. The large-scale correlations between halos are described using a linearly biased matter power spectrum (the two halo term), and small-scale correlations are described in terms of a radial halo profile (the one halo term). From Fig.~\ref{fig:voro}, it is clear that the Voronoi tessellation probes the intermediate regime, precisely outside of a halo's interior, but nearer to the halo than its nearest neighbors. Voronoi statistics may offer a way of testing and improving halo model calculations, precisely by providing information about these intermediate scales, but this is outside the scope of this work.
		
			\subsection{Stochasticity and nonlinearity}
			\label{ssec:dsv}
			
				For the matter PDF response, the two major challenges to providing an accurate model are due to the extreme nonlinearity of the scales it probes, and the stochasticity of Voronoi cell evolution on these scales. The typical approach to modeling a density PDF of the smoothed matter field relies on the fact that, at early enough times, the density statistics are very nearly Gaussian. Suppose we can provide a deterministic map from the late-time densities $\delta_\V(t_f)$ back to the early-time densities $\delta(t_i)$,
				\begin{align}
					F\big(1 + \delta_\V(t_f)\big) = \frac{\delta_\V(t_i)}{\sigma} \, ,
				\end{align}
				where $\sigma$ is the standard deviation of the early fluctuations. The quantity on the right-hand side of the above equation is drawn from a unit normal distribution. Then $\pdf\big(1 + \delta_\V(t_f)\big)$ is obtained by multiplying this unit normal distribution by the Jacobian of the mapping provided by $F$ \cite{Bernardeau:2001qr, Lam:2007qw},
				\begin{align}
					\pdf(1 + \delta_\V) = \frac{1}{\sqrt{2\pi}} \frac{\dee F}{\dee \delta_\V} \exp \lb -\frac{F^2}{2} \rb\, .
				\end{align} 
				
				As a first approximation, one might use the analytic solution for spherical collapse (expansion) of overdensities (underdensities) in the Einstein--de~Sitter cosmology. This mapping is plotted int Fig.~\ref{fig:icf} as the red dashed curves. While this approximation is an excellent starting point for modeling the Eulerian PDF, it cannot work for the Lagrangian PDF. For the Eulerian densities, the variance $\sigma^2$ is evaluated at an early-time length scale that encompasses the late-time mass within the smoothing window. For high-density perturbations, the length scale is large, and $\sigma$ is small on these large scales, so $F$ can take arbitrarily large values. 
				
				For the Voronoi density PDF, $\delta_\V$ is already smoothed on a fixed mass scale, so $\sigma$ is constant. The spherical collapse map, however, has a maximum $\delta_\V(t_i)$ corresponding to the critical density for collapse, so $F$ is bounded from above in this case, and cannot be described by a normal distribution. The problem can be seen in Fig.~\ref{fig:icf}, where a 2D contour plot of the map from late-time to early-time Voronoi densities is plotted for redshifts $z=0.0$ and $z=1.0$. We measured that Voronoi densities in our simulation initial conditions at redshift $z=49$, and matched them with the late-time Voronoi densities using the unique particle ID numbers.
	
				\begin{figure*}
					\centering
					\graphic{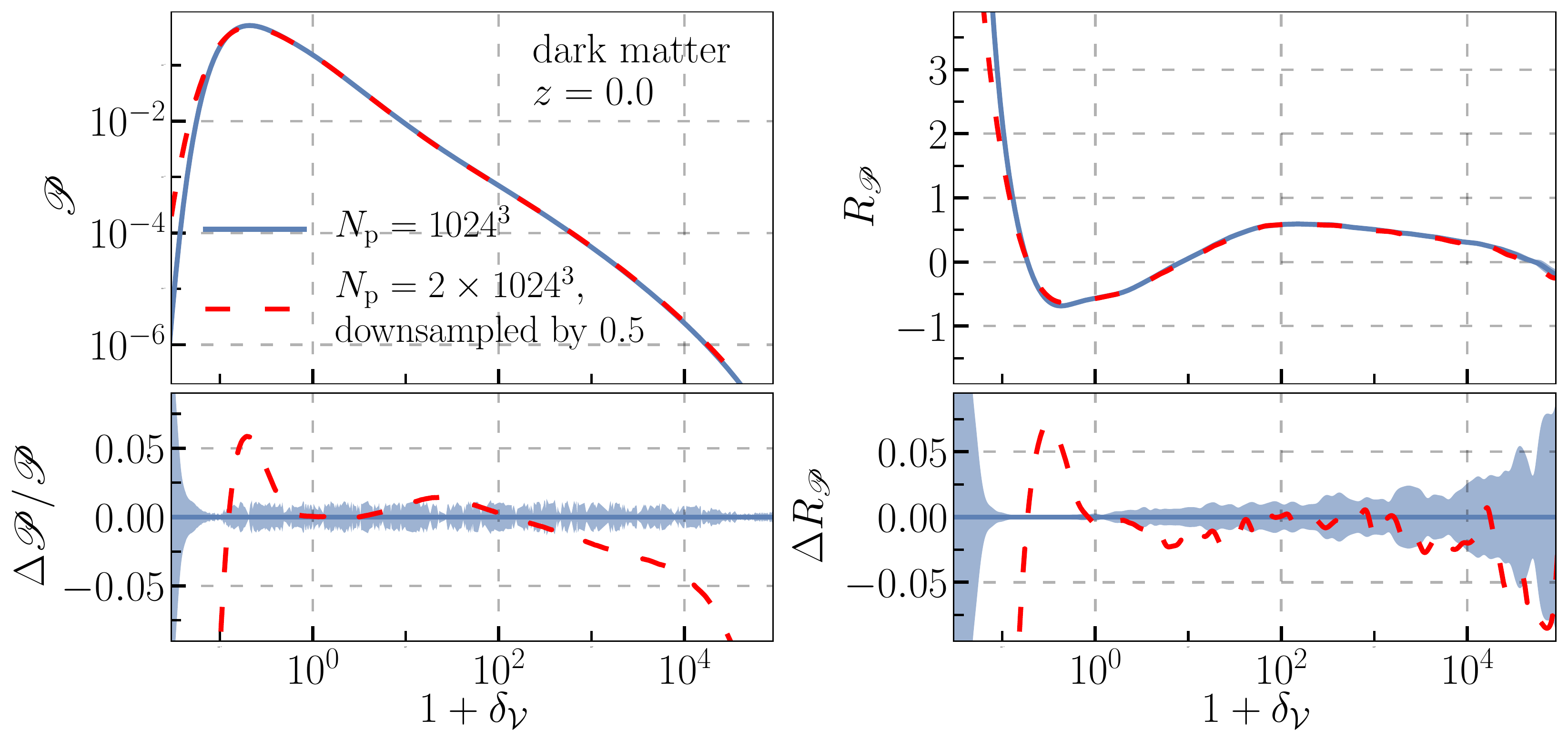}{1.}
					\caption{Resolution convergence test for the PDF (left) and its separate universe response (right). The fractional difference between the lower resolution and higher resolution PDFs is shown on the bottom left, and the absolute difference between the two responses is shown on the bottom right.}
					\label{fig:dsv}
				\end{figure*}
				
				Clearly the mapping is highly stochastic. Also, many particles are above the critical density for spherical collapse, which corresponds to the horizontal asymptote of the spherical collapse curve. While these high-density particles do end up within collapsed objects (i.e. halos), their collapse is eventually halted by virialization. The late-time virial densities are shown in Fig.~\ref{fig:icf} as the black dashed vertical lines.
				
				The dot-dashed curves of Fig.~\ref{fig:icf} are the median late-time densities at fixed early-time density. This quantity indicates what kind of environment a particle is likely to end up in if you know its early-time cell size. The solid blue curve shows the median early-time density at fixed late-time density. This quantity indicates which density a particle is likely to have started at if you know the size of its Voronoi cell at late times. Not only are the early densities are nearly Gaussian overall, but the distributions along vertical slices through the contour plots are also nearly Gaussian and thus symmetric about their peaks, so their mean, median, and mode coincide. The same is not true of horizontal slices through the contour plots. Along this direction, the mean curve is significantly to the left (towards late-time underdensity) of the median, and the mode is even further to the left, although we have only shown median curve.
				
				Interestingly, the spherical collapse curve roughly approximates the median early-time density curve below the virial density. This approximation is better at earlier redshifts. The crossing point of these two curves occurs near the virial density, which evolves with redshift. The crossing between the median and spherical collapse maps occurs at Voronoi densities $1+\delta_\V =$ 278, 181, and 140 at redshifts $z=1.0$, 0.5, and 0.0 respectively. The virial densities at these redshifts are $337\,\brho$, $203\,\brho$, and $148\,\brho$ respectively, so the crossing is nearer to the virial density at earlier redshifts. Above the virial density, the two curves diverge as the median early-time densities increase exponentially, and the spherical map asymptotes to the critical density for collapse. Note that many of the cells within the virial radius of a halo have Voronoi densities below the virial density of spherical collapse.
				
				At these extremely nonlinear scales, Lagrangian perturbation theory has been shown not to offer significant improvement with regards to this stochastic mapping \cite{Neyrinck:2012bf}. One way of introducing stochasticity into the spherical collapse calculation is to generalize it to ellipsoidal collapse \cite{Lam:2007qw}. This is an attractive approach, because the distribution of ellipsoidal regions is known analytically in a random Gaussian field \cite{1970Afz.....6..581D}. However, since the dark matter cells undergo dramatic flows into halos and their surrounding environments, and out of voids, a more detailed study of the dark matter velocity field is likely required to physically motivate an accurate model \cite{Jennings:2015mja}, and we leave this for future work.
							
			\subsection{Downsampling validation}
			\label{ssec:dsv}
			
				Since each Voronoi cell contains precisely one dark matter particle, each cell represents the limit of resolution achieved in the simulation. A reasonable concern is that the Voronoi statistics we measured in the dark matter could be dominated by finite-resolution artifacts. If that were the case, then we would have measured something that reveals more about the N-body simulations than it reveals about the physical matter density field that the simulations are intended to model.
				
				To verify that this is not the case, and to test how well converged the matter PDFs are, we ran a single set of separate universe simulations with twice the number of particles, $N_\mathrm{p} = 2\times (1024)^3$, and randomly downsampled to $1024^3$ particles. We downsampled on the particle ID numbers, so that the same particles were removed from the overdense, underdense, and global universe simulations. We then measured the separate universe response of the matter Voronoi density PDF from the higher resolution simulations.
				
				The results are plotted in Fig.~\ref{fig:dsv}. For cell densities in the range $-0.8 < \delta_\V < 10^4$, differences between the lower resolution and higher resolution PDFs are at most $5\%$, and the differences are significantly lower between $0 < \delta_\V < 10^3$. While the differences are clearly smooth, and thus systematic, they are a subdominant contribution to the overall shape of the distribution.
				
				For the separate universe response, the agreement between the different simulation resolutions is excellent for $\delta_\V > 0$. The oscillations in their difference indicate that the dominant contribution is due to statistical uncertainty. There also seems to be a small systematic error, causing the response from the lower resolution simulations to be 1--2\% above the higher resolution response. This suggests that much of the systematic error in the PDFs cancels in the separate universe response. For underdense cells, the response discrepancy is similar to the PDF discrepancy. Since the underdense cells, by definition, have the lowest particle-per-volume ratio, they represent regions of the simulation box where the worst resolution is achieved. This could explain why both the PDF and its separate universe response are less converged for underdense cells. This explanation is consistent with the fact that the underdense halo cells exhibit agreement between the separate universe and clustering methods, since isolated halos are less affected by changes in simulation resolution than isolated dark matter particles are.  
		
		\section{Conclusion}
		\label{sec:con}
		
			In this paper, we have measured the one-point statistics for the Lagrangian density field of matter in N-body simulations, approximated by the use of Voronoi tessellation (Fig.~\ref{fig:pdfs}). We also measured the one-point statistics for the Voronoi densities of simulation halos. Using separate universe simulations, we measured the linear response of the Voronoi density PDF to the presence of large-scale matter fluctuations (Fig.~\ref{fig:Rsu}). We verify that this linear response accurately predicts the cross-correlation of local PDFs in simulation subboxes with the matter density perturbations in those subboxes (Fig.~\ref{fig:suvc}). For the matter, this verification represents a test of the separate universe method over an unprecedented range of densities. We also tested the sensitivity of these simulations results to changes in resolution (Fig.~\ref{fig:dsv}).
			
			We interpreted the linear, separate universe responses of the matter PDF as the cosmic bias of regions of the density field, smoothed over a uniform mass scale, and selected based on their density. For the halos, we interpreted the separate universe PDF response as the bias of halos selected based on their Voronoi cell size. This can be thought of as a new kind of assembly bias. 
			
			By tracking the Voronoi volumes of simulation dark matter particles, we were able to determine the evolution of each individual Voronoi cell (Fig.~\ref{fig:icf}). The mapping from densities of Voronoi cells at late times back to their early-time densities is highly stochastic. However, we found the mapping for a median Voronoi cell with density below the virial density is qualitatively similar to the spherical collapse mapping. Further progress on modeling the global and position-dependent Voronoi density PDFs may required considering aspherical dynamics, and studying the bulk flows of matter into halos and out of cosmic voids.
			
			Observationally, it may be possible to extract an analogous observable to the matter Voronoi density PDF from weak lensing maps. For example, this could be achieved by partitioning a lensing map into regions with a fixed mean amount of lensing convergence. The one-point distribution of the sizes, or densities, of these regions could then be computed. For galaxy surveys, a catalog of observed galaxies can easily be Voronoi tessellated directly. However, extracting cosmological information from galaxy tessellations is limited by redshift errors, and possibly redshift space distortions \cite{Cooper:2005ci}. The former issue may be dealt with through advanced statistical methods for improving observational redshifts, such as hierarchical modeling \cite{Jasche:2011qm, Malz:2020epd}. The effects of redshift space distortions need to be determined in simulations or mock observational data sets.
			
			Our investigation was limited to adiabatic perturbations in the standard $\lcdm$ cosmology. In this context, long-wavelength density perturbations undergo a universal growth, which is independent of scale in the linear regime. It has been previously demonstrated that extensions to $\lcdm$, including dynamical dark energy \cite{Chiang_2016, Jamieson_2018}, massive neutrinos \cite{Chiang_2018}, as well as primordial non-Gaussianity \cite{Dalal:2007cu}, lead to scale-dependent effects on large scales. Such effects would also appear in the shapes of PDF responses considered here, and cause them to become scale dependent on large scales. The Voronoi density PDF and its linear response may be a useful observable for targeting specific regions or features of the density field that are sensitive to these scale-dependent effects.

			\begin{figure*}
				\centering
				\graphic{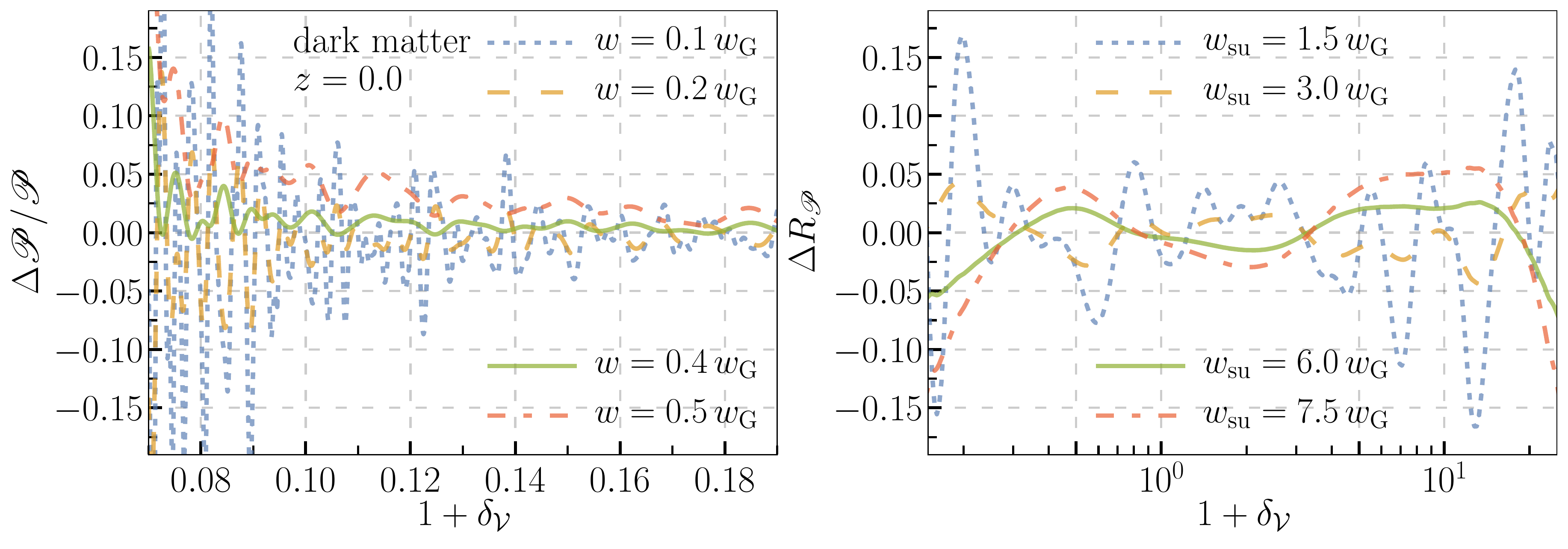}{1.0}
				\caption{Comparison of the halo Voronoi density PDF (left) and its separate universe response (right) when computed with different KDE smoothing widths. The values chosen for our reported results are $w = 0.3 \, w_{\mathrm{G}}$ and $w_\su = 4.5 \,w_{\mathrm{G}}$ and the differences above are computed with respect to these reference widths. From the plots, choosing a narrower width yields results that oscillate around those obtained with our chosen widths, demonstrating they have higher variance but that our results are not statistically biased with respect to them. Choosing a wider width yields statistically biased results, which the narrower width results do not oscillate around.} 
				\label{fig:kdew}
			\end{figure*}
			
			\acknowledgements 
			
			We would like to thank Matias Zaldarriaga and Ben Wandelt for useful discussions concerning this work. Results in this paper were obtained using the high-performance computing system at the Institute for Advanced Computational Science at Stony Brook University. Our figures were made using the P{\sc ython} package M{\sc atplotlib} \cite{Hunter:2007}, and many of our calculation were carried out using S{\sc ci}P{\sc y} \cite{Virtanen_2020} and N{\sc um}P{\sc y} \cite{Harris_2020}. D.J. is supported by Grants No. NSF PHY-1620628 and DOE DE-SC0017848. M.L. is supported by Grant No. DOE DE-SC0017848.
			
	\appendix*
	
	\section{KDE width selection for the PDF and its separate universe response}
	\label{sec:app}

	The choice of KDE width is important because the level of smoothing controls the variance and statistical bias of the estimator. A width that is too narrow results in excess variance, while a width that is too wide results in a highly biased estimator. According to Eq.~(\ref{eq:kdeb}), the bias of $\pdf_{\log}$ is proportional to the second derivative of the distribution. Converting to $\pdf$ using Eq.~(\ref{eq:plog}) contributes bias proportional to the first derivative. Since the slope of the PDF is very steep in the low-density tail, this part of the PDF is not as reliably estimated compared with higher densities.
	
	In Fig.~\ref{fig:kdew}, we demonstrate how the optimal kernel width can be chosen simply by inspecting results obtained using different widths. We first estimated the PDF in the global cosmology with a KDE width that is too narrow, $w=0.1\, w_{\mathrm{G}}$, which results in excess variance but negligible bias. We then recomputed the KDE using a wider smoothing width. If the KDE with a broader kernel width is not too biased, the fractional difference between the two estimators should oscillated around zero. In Fig.~\ref{fig:kdew}, we can clearly see that this is the case for the halo PDF using $w=0.3\, w_{\mathrm{G}}$. Repeating this process with wider kernel widths, eventually the bias of the estimator becomes noticeable. For example, with $w=0.4\, w_{\mathrm{G}}$, the bias is around 3\% at $\delta_\V = -0.93$, and with $w=0.5\, w_{\mathrm{G}}$ the bias is 15\% at the same Voronoi density. If we computed the fractional difference between the PDFs estimated using these wider kernels and the PDFs estimated using the narrower kernel widths, they would not oscillate around zero. The goal is to choose the widest kernel without any significant estimator bias. We only show the results at low density in Fig.~\ref{fig:kdew}, since that is where the estimator bias is most severe.
	
	We have shown the same method for kernel width selection for the separate universe PDF response in the right panel of Fig.~\ref{fig:kdew}. Here, we used $w=4.5\, w_{\mathrm{G}}$ as our reference. The narrower KDE widths oscillate around the reference response, while the broader widths yield a noticeable estimator bias. In this case, we have computed the absolute difference instead of the fractional difference, since the response has zero crossings. The effects of statistical bias are much milder for the response, allowing for a much wider smoothing width. This is due to the partial cancellation of the estimator bias when subtracting the overdense and underdense separate universe PDFs. 
	
	For the PDF clustering biases presented in Sec. \ref{ssec:guv}, the cancellation of oversmoothing effects is not as good. To optimize the kernel widths for the subboxes, we relied on Eq.~(\ref{eq:loc}). The global universe PDF's slope, which appears in the background shift term of Eq.~(\ref{eq:loc}), benefits from a similar level of estimator bias cancellation as the separate universe response. The optimal smoothing of the subbox data can then be chosen by computing the clustering for both $\pdf_{\mathrm{loc}}$ and  $\pdf_\su$, and selecting the widest width for which Eq.~(\ref{eq:loc}) is satisfied within statistical error. We found a narrower width of $w_\su = 2\, w_\mathrm{G}$ is required to satisfy this condition, which causes excess variance in the clustering method results.

	\bibliography{PositionDepedendentVoronoiPDF.bib}
		
\end{document}